\documentclass{article}
\usepackage{emulateapj,apjfonts}

\input psfig

\begin{document}

\def\kms{$\rm {km}~\rm s^{-1}$}
\def\ni{\noindent}
\def\msun{{\rm M}_\odot}
\def\deg{\ifmmode^\circ\else$^\circ$\fi}
\def\etal{{\it et~al.}}
\def\ie{i.e.,}
\def\eg{e.g.,}

\title{CFHT Adaptive Optics Observations of the Central Kinematics in M15}
\author{Karl Gebhardt\altaffilmark{1}\altaffilmark{,2}} 
\affil{University of California Observatories/Lick Observatory\\ 
University of California, Santa Cruz, Ca 95064}
\affil{gebhardt@ucolick.org} 
\author{Carlton Pryor\altaffilmark{1}, R. D. O'Connell\altaffilmark{1}, 
T.B. Williams}
\affil{Department of Physics and Astronomy, Rutgers, The State
University of New Jersey\\ 136 Frelinghuysen Rd., Piscataway, NJ 08854-8019}
\affil{pryor@physics.rutgers.edu,williams@physics.rutgers.edu}
\author{James E. Hesser\altaffilmark{1}}
\affil{Dominion Astrophysical Observatory\\ Herzberg Institute of
Astrophysics, National Research Council of Canada\\
5071 W. Saanich Road, R.R.5, Victoria, B.C., V8X 4M6, Canada}
\affil{James.Hesser@hia.nrc.ca}
\altaffiltext{1}{Visiting Astronomer, Canada-France-Hawaii Telescope,
operated by the National Research Council of Canada, the Centre
National de la Recherche Scientifique of France, and the University of
Hawaii.}
\altaffiltext{2}{Hubble Fellow.}

\begin{abstract}

We have used an Imaging Fabry-Perot Spectrophotometer with the
Adaptive Optics Bonnette on the Canada-France-Hawaii Telescope to
measure stellar radial velocities in the globular cluster M15
(NGC~7078).  An average seeing of 0.15\arcsec\ full-width at half
maximum, with the best-seeing image having 0.09\arcsec, allowed us to
measure accurately the velocities for five stars within 1\arcsec\ of
the center of M15.

Our estimate of the second moment of the velocity distribution (\ie\
the dispersion, ignoring rotation) inside a radius of 2\arcsec\ is
11.5~\kms, the same value we find out to a radius of about 6\arcsec.
However, the projected net rotation does increase dramatically at
small radii, as our previous observations led us to suspect.  The
rotation amplitude inside a radius of 3.4\arcsec\ is $v =
10.4\pm2.7$~\kms\ and the dispersion after removing the rotation is
$\sigma = 10.3\pm1.4$~\kms, so $v/\sigma \simeq 1$ in this region.
In addition, the position angle (PA) of the projected rotation axis
differs by 100\deg\ from that of the net cluster rotation at larger
radii. Current theoretical models do not predict either this large an
increase in the rotation amplitude or such a change in the PA.
However, a central mass concentration, such as a black hole, could
possibly sustain such a configuration. The rotation increase is
consistent with the existence of a central dark mass concentration
equal to 2500~$\msun$.

The Strehl ratio is 1\% in our worst images and 6\% in our best.
Despite these low values, the images allow us to resolve the brighter
stars with an angular resolution close to the diffraction limit and to
perform photometry on these stars accurate to a few percent.  Thus,
these adaptive optics observations provide us with crucial information
on the central kinematics of M15.

\end{abstract}


\section{INTRODUCTION}

M15 (NGC~7078) was one of the first globular clusters known to have a
steep central stellar density profile, inconsistent with isothermal
models (King 1975, Djorgovski~\& King 1986).  Its central structure is
usually interpreted to be a product of ``core-collapse'' (\eg\
Grabhorn~\etal\ 1992)---a gravo-thermal instability that leads to a
very small and dense core surrounded by a power-law cusp.
Core-collapse predicts a specific density profile, about $r^{-2.2}$,
and velocity dispersion profile, about $r^{-0.1}$ or nearly constant,
in the cusp for the stellar component that contributes most of the
mass (Cohn 1980, Spitzer 1987). However, the observed luminous
component may have different profiles because of mass segregation.
Thus, both photometric and kinematic observations are needed to
estimate the profile of the gravitational potential and test the
models.

The stellar density profile near the center has been measured by many
authors; the best study comes from the HST analysis of
Guhathakurta~\etal\ (1996), who show that M15 contains a central cusp
with a logarithmic slope around --1.8.  This estimate is based on the
deprojection of individual positions of the central stars.  Although
M15 has one of the highest surface brightnesses among the
centrally-concentrated clusters, at the radii of interest these
measurements still suffer from poor statistics due to the small number
of stars.  Thus, the uncertainty of the slope within a radius of
1\arcsec\ is around 50\%.  Even more uncertain is the two-dimensional
distribution of stars; \ie\ whether the central isophotes are
spherical or flattened.

Unfortunately, kinematical profiles are more difficult to measure than
the light distribution and suffer from larger uncertainties.  The
central kinematics of M15 have been the subject of numerous studies
(Newell~\etal\ 1976, Peterson~\etal\ 1989, Dubath~\etal\ 1995,
Zaggia~\etal\ 1993, Gebhardt~\etal\ 1994, 1995, 1997, Dull \etal\
1997, Drukier \etal\ 1998).  Measuring the
central dispersion has been the primary goal in most of these studies.
However, the central stellar density is so high that severe blending
of the stellar images limits the number of stars that can be measured
individually from the ground, while the large range of stellar
luminosities on the giant branch limits the number of stars making
significant contributions to the integrated light.

The improvements in our understanding of the central region have come
from data taken under extremely good conditions (\ie\ Guhathakurta's
study of the stellar distribution using HST) or from the use of
instruments designed to optimize the seeing profiles (\ie\
Gebhardt~\etal's 1997 study using the fast-guiding Sub-arcsecond
Imaging Spectrograph on the Canada-France-Hawaii Telescope (CFHT)).
However, we have not achieved the optimal situation of equal angular
resolution for the spatial and kinematic observations, primarily
because the relatively small aperture of the HST makes spectroscopy
very expensive.  Combining adaptive optics (AO) observations, which
yield the highest spatial resolution possible from the ground, with
the two-dimensional velocity reconstruction enabled by Fabry-Perot
(FP) observations should provide the most accurate measurements for
the central kinematics.

Adaptive optics is one of the most powerful tools available to
astronomers today, potentially allowing imaging at the diffraction
limit of a ground-based telescope.  AO is presently most efficient at
near-infrared wavelengths.  However, high-resolution spectrographs are
only beginning to be used there, so most kinematic studies must still
rely on optical wavelengths.  Fortunately, in an overlap region near
9000 \AA, spectrographs and CCDs are efficient, while AO systems still
yield significant corrections.  However, in this region the Strehl
ratio -- the ratio of the central intensity in the real stellar
profile to that of a diffraction-limited profile -- generally lies
below 20\%.  This ratio is one of the most important numbers
characterizing the quality of the corrected images.  Values that low
mean that careful measurements of the stellar point spread function
(PSF) are required to interpret the results.

Globular clusters provide an ideal target for AO observations because
the field is filled with stars that yield the PSF.  Slit or aperture
spectrographs do not lend themselves to AO kinematic studies because
they do not generally allow an accurate measurement of the spatial PSF
and because it can be hard to determine exactly where the slit was
placed.  Fabry-Perot imaging, in contrast, does provide both accurate
spectral and accurate spatial information.  In this paper we describe
results from using the Adaptive Optics Bonnette (AOB) on the CFHT with
an imaging Fabry-Perot.

The results from these data provide crucial information for the region
within 3\arcsec\ of the center of M15. They confirm the earlier
results of Gebhardt~\etal\ (1997) that the observed velocity
dispersion profile remains flat inside a radius of a few arcseconds
and that the net projected rotation profile rises in that region.  The
increase in the rotation is difficult to understand, but might be
explained by the existence of a central massive dark object.

The rest of this paper is organized as follows.  Section~2 describes
how the FP AO data were taken and reduced, including a discussion of
the impact that the small and variable Strehl ratio had on the
accuracy of our photometry.  Section~3 presents the profiles of the
projected velocity dispersion and net projected rotation that result
from these data, while Sec.~4 briefly revisits the mass density
profiles derived with the techniques of Gebhardt~\etal\ (1997).
Finally, Sec.~5 summarizes the results and discusses how the large
rotation at small radii in M15 might be understood in the context of
the dynamical models of globular cluster evolution.

\vskip 20pt
\section{THE DATA}

\subsection{Instrumentation}

In normal use, a Fabry-Perot \'etalon is placed in a collimated beam to
avoid degradation of the spectral resolution.  With the AO system on
the CFHT at the time of our observations (see the description in
Rigaut~\etal\ 1998), this placement was not possible.  In consultation
with the CFHT staff, we chose to place our etalon in the f/40
converging beam behind the AOB focal expander.  The STIS2 CCD then
yielded 0.0305\arcsec\ pixels and a 63\arcsec\ field of view.
However, the AOB optics are not telecentric -- the optical axes of the
converging beams were not perpendicular to the etalon away from the
center of the field -- and this caused additional loss of wavelength
resolution.  The converging beam caused a negligible degradation of
the 2.0~\AA\ intrinsic resolution of the etalon at the center of the
field, but by a radius of about 6\arcsec\ the resolution was worse by
a factor of 1.4.  Because our primary targets were the centers of
clusters, this limitation of the usable field did not significantly
compromise our scientific goals.

Our usual observing procedure is to scan across the H$\alpha$ line,
since it is a deep line that is not very sensitive to metalicity.
With AO, we cannot obtain adequate correction at wavelengths that
short.  Fortunately, the coatings of the Rutgers etalon allowed us to
work with the 8542~\AA\ CaII triplet line.  We employed our own filter
to isolate the correct etalon order.  It had a passband with a 30~\AA\
full-width at half maximum (FWHM), centered on 8542~\AA.  A dichroic
in the AO system sent the I-band light to the etalon and the CCD, and
all other wavelengths to the wavefront sensor.

\subsection{Observations and Data Reduction}

We took data on June 15--19, 1998, observing M15 on three of the nights
and obtaining a total of 26 usable 15-minute exposures for this cluster
stepped across $\lambda\lambda 8536$--8543~\AA.  We observed three other
clusters (M13, M30, and M80) as well, which we will present in a later
paper.

Due to the converging and non-telecentric beam, we could not use our
standard wavelength calibration procedures and the result was increased
uncertainty in the solution.  Normally, a single exposure of a screen
illuminated by an emission-line lamp produces a bright ring in the
image because of the simple quadratic dependence of transmitted
wavelength on distance from the optical axis.  We measure the radii of
these rings to calibrate the relation between etalon spacing and
wavelength.  However, in the present case, measurable rings did not
exist due to a complicated dependence of both the mean wavelength and
the spectral resolution on the distance from the optical axis.  We were
thus forced to take a sequence of images that scanned the etalon across
the calibration line to measure the etalon spacing corresponding to the
peak of the line.  This procedure produced an adequate wavelength
calibration (uncertainties $<0.1$~\AA), but required more time than
usual.

We corrected for pixel-to-pixel sensitivity variations and the change in
the transmission of the order-separating filter with wavelength using
images of an internal incandescent flat field.

The CFHT AO system requires a natural guide star in the field to
correct the wavefront.  For the globular clusters studied, the choice
was governed by a tradeoff between brightness and distance from the
center.  The central brightness cusps of M15 and M30 had too much
structure to act as good guide stars.  For M15, we used the star AC3
(Auri\`ere~\& Cordoni 1981), which has $V = 13.3$ and is 6.7\arcsec\
away from the center.  The V-band magnitudes of the guide stars ranged
from 13.3 to 15.5, with the best correction coming from the brighter
stars.

\subsection{Photometry with AO}

The high stellar density near the center of M15 produces crowded
frames that force us to use profile-fitting photometry as opposed to
aperture photometry.  Thus, errors in estimating the PSF directly
affect the photometry and accurate photometry depends on understanding
the PSF both as a function of position on the chip and as a function
of time throughout the observations.  AO in the optical complicates
both of these matters; the small isoplanatic patch causes the PSF to
vary significantly over the field and, since the AO correction is
highly dependent on the native seeing, sudden changes in the seeing
cause large temporal variability in the PSF.  Some published reports
suggest that these problems will limit the accuracy of stellar
photometry in crowded fields with AO to be no better than 10\%
(Roberts \etal\ 1997, Esslinger \& Edmunds 1998).

Estimating the PSF is a hard problem because the profile for a stellar
source observed with the AO system is approximately a combination of
an unguided profile -- the natural site seeing -- and the
diffraction-limited profile, so the low Strehl ratios in the optical
imply that most of the stellar light resides in the uncorrected PSF
and very little in the diffraction-limited core. The fraction of the
flux in the diffraction-limited core is approximately equal to the
Strehl ratio.  Therefore, we need to use very large PSFs to accurately
measure all of the light at large radii, but, at the same time, we
must finely sample the PSF to maintain the high-spatial-resolution
information provided in the diffraction-limited core.  Getting enough
signal-to-noise to measure accurately the wings of the PSF is
difficult.

Figure~1 shows Strehl ratios for the guide star in each M15 frame
plotted as a function of FWHM of the star.  The Strehl ratios range
from 0.01 to 0.06; for our typical exposure over 90\% of the light is
in the broad component of the PSF.  Although these Strehl ratios are
low, accurate photometry depends {\it solely} on the ability to
measure the PSF.  Fortunately, the centers of globular clusters have
bright stars scattered throughout the field, and so are well suited to
measuring the PSF well into its wings with high-spatial resolution.

Because with FP observations we are building a spectrum from data that
was taken over several days, the accuracy with which we measure the PSF
in each individual frame is one of the most crucial aspects of the
whole reduction.  Previously, in these crowded fields, we have obtained
photometry errors of roughly 2-3\%; the errors are
distributed among the following sources: crowding, the wings of the
PSF, frame-to-frame normalizations for changing transparency, and
wavelength calibration.  With adaptive optics, we have to pay special
attention to the PSF measurement.  Photometry errors greater than
10\%\ would render the data useless, since such errors would result in
velocity uncertainties larger than 10~\kms, an unusable uncertainty for
most globular clusters.  We discuss below in turn the spatial extent of
the PSF, the spatial variability of the PSF within a frame, and the
temporal variability between frames.

\subsubsection{PSF Extent}

We determined the PSF for each frame with the normal DAOPHOT routines
(Stetson 1994) and obtained the final photometry using ALLFRAME
(Stetson~\etal\ 1998). A quadratic variation with position on the chip
provided an excellent fit for the PSF across the frame. We paid
particular attention to the extent to which the PSF was both fitted
and subtracted. For our observations, the diffraction limited core of
the PSF is about 0.05\arcsec\ across and the broader component is
about 0.5\arcsec. This means that we must use a PSF that goes out
beyond 1\arcsec\ to contain most of the light.  Given the pixel scale,
we used a PSF radius of 50~pixels and a fitting radius of
20~pixels. The large PSF radius mandates using a large number of stars
to overcome the low signal-to-noise in the wings of the PSF.  Fig.~2
demonstrates the PSF's extent by plotting the flux of the guide star
in each M15 frame versus radius. We truncate the logarithmic plot at a
radius of 60~pixels (1.8\arcsec), but light is still present at these
radii and our ALLFRAME photometry thus requires aperture corrections.

We can verify this extent for the PSF from data for an isolated
standard star. Four consecutive 10s exposures of HD107328 (HR 4695;
V=4.96) were taken in twilight on the fourth night.  Conditions were
mildly non-photometric; the Strehl ratios were about 0.03.  The filled
circles in Fig.~3 show the RMS scatter around the average aperture
magnitude for apertures with radii between 3 and 150 pixels.  The
triangles are the uncertainties in the magnitudes expected from photon
statistics.  The RMS scatter in the magnitudes decreases with
increasing aperture size until a radius of 50 pixels, after which it
slowly increases.  This scatter suggests that changes in the PSF
caused by variations in the seeing and wavefront correction have
become small at an aperture radius of 50 pixels.  The amount of light
within this radius is about 87\% of that within a radius of 150 pixels
for these data.

Problems caused by the extended PSF are shown in Fig.~4.  This figure
contains four images: (1) the best-seeing CFHT frame displayed with a
linear stretch, (2) an HST frame of the same field as (1), (3) a
slightly larger region of the same CFHT frame displayed with a
square-root stretch, and (4) the same field as (3) with the stars
subtracted.  The CFHT image has a smaller FWHM than the HST image;
however, the later image is preferable when one takes the tails of the
PSF into consideration. The HST image essentially has no tails,
whereas for the AO image most of the light is in the
tails.\footnote{In this regard our AO images are more like pre-COSTAR
HST images.}  The bottom-left image demonstrates the amount of light
in the tails since, with this stretch, stars separated by 1\arcsec\
start to blend together.  The bottom-right, star-subtracted, image
shows residual light where there is a high density of stars because
our estimate of the PSF does not contain all of the light. Since this
light correlates with the brightness of the subtracted stars, it
cannot be unresolved stellar light and is instead light from the
unmodeled outer fringes of the PSF.

\subsubsection{PSF Variability}

We show the spatial dependence of the PSF in Figs.~5 and 6.  Figure~5
overplots radial profiles of four PSFs at distances of 0, 10, 20, and
30\arcsec\ from the guide star.  The PSFs are normalized to have
identical volumes.  For each profile, the broad component -- the
uncorrected profile -- is essentially unchanged, as we expect. What
does change is the amount of light in the diffraction-limited core, as
stars at the largest radii have almost no light in the core due to the
AO correction.  To demonstrate this effect with the FWHM, each line in
the top panel of Fig.~6 shows the ratio of the FWHM of the PSF to the
FWHM of the guide star as a function of CCD column number for the
central row of an M15 frame.  The ratio changes by a factor of two from
the middle of the image to its edges.  In addition, the change is
similar for each frame.  There is a slight dependence on the base FWHM;
the bottom plot of Fig.~6 presents the FWHM ratio at the edge of each
image as a function of the base FWHM.  The smallest FWHM image has the
largest variation across the field.

Figure~7 shows the temporal variation of the PSF by plotting the FWHM of
the guide star as a function of frame number.  The vertical lines mark
the divisions between the three nights.  Large, but smoothly varying,
changes in the FWHM occur throughout the night.  The greatest variation
occurs for exposures during morning twilight, where the additional
background begins to affect the AO corrections.

\subsubsection{Photometric Accuracy}

The ideal situation for determining the photometric accuracy of the
observations would be to expose frame after frame using an identical
setup.  We obviously cannot employ this ideal method since we must scan
across the absorption line in order to measure stellar velocities, but
we can use our measured absorption line profile to estimate photometric
accuracy.  Fig.~8 plots the line profiles for four bright stars in our
field.  The line is a fitted Voigt profile (see Gebhardt \etal\ 1994
for details).  The deviation of the points from the fitted line result
from the photometric errors; the error bars for each point come from
the standard DAOPHOT uncertainties.  If the actual deviations divided
by the expected uncertainties do not have a Gaussian distribution with
unit standard deviation, the additional uncertainty is probably caused
by our lack of understanding of the PSF.  The points to which we fit
Voigt profiles, such as those in Fig.~8, are already corrected on a
frame-by-frame basis for the combined effects of transparency
variations and the truncation of the PSF.  As discussed in the
beginning of Sec.~2.3, this last effect arises because our measured PSF
does not include all of the light.  Since we have many stars at
different radial and spectral (\ie\ different velocities) positions,
these frame normalizations are unique and well-measured (see Gebhardt
\etal\ 1994 for more discussion).

We average the fractional deviations from the fitted line profile for
every star in each frame in order to estimate the true photometric
uncertainties for each of the frames. Each point in Fig.~9 is the rms
dispersion of the fractional differences between the fitted profiles
and the data points in a frame as a function of the FWHM of the guide
star in the frame. We estimate the dispersion from the biweight
estimate of scale (Beers, Flynn, \& Gebhardt 1990) using all stars
with a velocity measurement and breaking the sample of stars into
three groups according to brightness.

The photometric uncertainties indicated in Fig.~9 result from both
photon statistics and any uncertainties due to particulars of the
analysis---\ie\ difficulties in measuring the PSF and the wavelength
calibration. DAOPHOT provides an estimate of the photon noise through
its reported uncertainty. For the brightest stars (the filled circles
in Fig.~9), the typical DAOPHOT uncertainty is around 0.5\% and, for
the faintest stars, it varies from 6 to 10\% (depending on the PSF
FWHM). These typical uncertainties are smaller than the actual
photometric scatter shown in Fig.~9 for both the faint and the bright
stars. This difference does not suggests an additive uncertainty
independent of brightness, such as that expected from PSF errors, but
rather a multiplicative error in the DAOPHOT uncertainties---they are
about a factor of 1.5 too small. Making such a correction brings the
observed and predicted uncertainties of the faint and medium stars
into reasonable agreement. However, the dispersion around the profiles
of the bright stars shown in Fig.~9 is still larger than
expected. This trend suggests the presence of an additive uncertainty
of about 2\%. We take this value as the upper limit to errors in our
photometry caused by the difficulty of estimating the AO-corrected
PSF.

\subsection{Velocity Zero-Point and Uncertainties}

In all of our Fabry-Perot runs we correct the zero-point of our
velocities since it is difficult to get an accurate absolute wavelength
calibration -- we build up the spectrum over the course of a few nights
and each frame has to be accurately calibrated.  As discussed in
Sec.~2.2, our current wavelength calibration is less accurate than for
previous observations and, for this reason, we have an increased
velocity zero-point uncertainty.  However, we do not believe that we
have systematic errors in our velocities as a function of velocity
because we did not observe adjacent wavelength points at adjacent
times.

To determine the velocity zero-point, we compare our velocities to the
many previous datasets for M15: Gebhardt \etal\ (1994, 1997), Peterson
\etal\ (1989), Dull \etal\ (1997), Drukier \etal\ (1998), and Dubath
\& Meylan (1994). The present data have been compared to each of these
individually and to a combined dataset.  All of the comparisons yield
the same result: the required velocity offset is 5.7~\kms.  In
addition, one must increase the uncertainty for each AO velocity by
adding 3.0~\kms\ in quadrature with the estimated uncertainty.
Previous FP studies added 0.5--1.0~\kms\ in quadrature, which is
consistent with the intrinsic velocity jitter of the giant stars
(Gunn~\& Griffin 1979, Mayor~\etal\ 1983).  We attribute the larger
errors in the present study to the difficulty of the wavelength
calibration.  Fig.~10 plots the difference between the 1995 CFHT/SIS
and the shifted AO velocities as a function of the 1995 velocities.
The two sets of data agree within their uncertainties, and give a
reduced $\chi^2 = 1.2$ for 67 degrees of freedom.

We list the mean radial velocities in Table~1. For each star, we give
the ID (Col.~1), the offsets in right ascension and declination
(Cols.~2 and 3) from the center (as defined in Guhathakurta~\etal\
1996) in arc-minutes, the velocity and its uncertainty (Col.~4 and 5),
the V-band magnitude from either HST or ground-based photometry
(Col.~6), the probability (Col.~7) of the $\chi^2$ from the multiple
measurements exceeding the observed value by chance (blank if there is
only a single measurement), and the ID (Col.~8) from Auri\`ere~\&
Cordoni (1981). The ID in Col.~1 is that of the HST WFPC2 photometry
of Guhathakurta~\etal\ (2000).  When those did not exist we generated
our own sequence numbers starting with 40001. The reported velocity is
the classical mean of all of the measurements for that star, each
weighted by the inverse of the square of the measurement uncertainty.
The uncertainty in the mean is derived classically as well and will
not be a good estimate of the true uncertainty when the $\chi^2$
probability is low.  Table~2 lists the individual velocity
measurements from the eight epochs of data.  The zero-point correction
discussed above has been applied to the AO velocities, so all of the
sets are consistent with the zero-point of PSC (we note that the
Fabry-Perot velocities from 1991 to 1994 are the same as in
Gebhardt~\etal\ 1997; however, the 1995 FP velocities are 0.6~\kms\
larger due to a better estimate of the zero-point). There was no
zero-point correction for either the Drukier or the Dubath data. In
contrast, the additional uncertainties that we have adopted ---
3~\kms\ for the present dataset, 1.5~\kms\ for FP95, 2.0~\kms\ for
FP94, 1.5~\kms\ for FP92, 1.5~\kms\ for FP91, 0.7~\kms\ for PSC,
0.7~\kms\ for Drukier, and 0.7~\kms\ for Dubath --- have not been
included in the listed uncertainties.  These additional uncertainties
were used to calculate the velocity uncertainties and the $\chi^2$
probabilities given in Table~1.

In Gebhardt~\etal\ (1997), there is an inconsistency between the IDs
of the stars in their Table~1 compared to those in their Table~2.  For
this reason we do not repeat those IDs in Table~1 here, instead
introduce new IDs which are consistent between the two tables
presented in this paper.  Since all of the velocity data is provided
in Table~2, there is no need for a cross-reference with previous
papers.

\section{RESULTS}

The total velocity sample for M15 yields 1773 stars with a mean
cluster velocity of --$107.5\pm 0.2$~\kms.  Only the sampling
uncertainty is included in the standard deviation.  The AO dataset
contains 104 stars with velocity uncertainties smaller than 10~\kms.
Keeping only the best data based on visual inspection of the profiles
provides a sample of 82 stars, but does not significantly alter the
results; we thus use the full dataset in the following dynamical
analysis. Given a total sample of 1773 stellar velocities, the 104
stars measured here only provide very modest gains in the total sample
size.  However, we stress that these measurements are in the central
regions, where it is crucial to determine the kinematics accurately,
as these regions are of the most importance for measuring the current
dynamical state of the cluster. 

In the central 1.5\arcsec\ radius, there are five stars (the first
five listed in Table~1) with velocities that the present dataset
measured more accurately than previous datasets. The line profiles
from the AO dataset for four of the five stars are shown in Fig.~11.
One of these five stars, AC215 (number 5933 in Table~1 and Fig.~11),
is a potential binary based on three epochs of observations; however,
a detailed binary analysis will be the subject of a future
paper. Another, AC214 (number 5831 in Table~1 and Fig.~11), is at
least three stars (HST IDs 5831, 5846, and 5872) according to the
photometry of Yanny~\etal\ (1994) and Guhathakurta~\etal\ (1996).
These authors point out that this clump of stars is a candidate for
the center of the cluster and note that the velocities of the stars
should differ by $\geq$40~\kms\ if the center contains
$\geq$$10^3~\msun$ concentrated within the approximately 1000~AU
projected extent of the clump.  Since we have not separated AC214 into
three components, we could only measure these velocity differences by
a broadening of the average line profile.  Our fitted profile for this
object is not larger than what we measure for the other stars in the
sample.  However, one of the three stars, ID~5831, is 0.8 and 2.0
magnitudes brighter in the I~band than the other two.  It is also the
reddest, hence the most strong-lined, of the three (which lie on the
horizontal branch or its transition to the asymptotic giant branch;
Guhathakurta~\etal\ 1996), so it is likely that our measured profile
mostly reflects that of this star.  In any case, the signal-to-noise
of the spectrum is very poor---likely due to the PSF fitting being
confused by the three components and the variable seeing---making any
firm conclusions impossible.

With only a small number of new velocity measurements compared to the
previous study by Gebhardt \etal\ (1997), we do not present a complete
dynamical analysis of M15.  Since, however, all of the velocities in
this study are in the central 20\arcsec, we present the kinematic
state of the central regions of M15.  The previous dataset for M15
contained 15 stars inside a radius of 2.5\arcsec. The present data
adds eight new stars, and confirms the velocity measurements for most
of the original 15 stars.  With such small samples, it is important to
have multiple measurements to secure the velocities. Two competing
complications explain why we were not able to measure velocities for
all of the stars in the previous samples: crowding and stellar flux.
The AO system does allow for better separation of crowded stars;
however, it requires more light due to the greater number of pixels
over which a star is spread. Thus, our previous non-AO observations
achieved fainter flux limits, whereas the AO system provided better
separation in the most crowded regions.  In total, there are 34 new
velocity measurements from the AO system, all for stars at radii less
than 17\arcsec.

\subsection{Kinematics Inside 20\arcsec}

Our goal here is to measure the moments of the velocity distribution as
functions of radius to constrain the dynamical state of M15. The second
moment about the mean cluster velocity represents a straight-forward
quantity that can be directly compared to dynamical models. A more
difficult measurement is determining the amplitude of the projected net
rotation ($v$) and projected velocity dispersion ($\sigma$) separately;
it is especially difficult for systems such as globular clusters, where
$v/\sigma$ is small. We estimate each of these quantities in turn
below.

Fig.~12 plots the radial profile for the second moment of the velocity
distribution for M15. The solid and dashed lines represent an estimate
and its associated 90\% confidence band obtained using the LOWESS
technique described in Gebhardt~\etal\ (1994).  At around 2\arcmin,
the density of points decreases drastically as we shift from
Fabry-Perot data to other datasets.  Due to the change in density, our
LOWESS technique artificially biases the second moment in the lower
density regions towards that in the higher density regions. This bias
results from our windowing function, which includes a set number of
data points rather than a specified radial range.  For that reason, we
do not use the LOWESS estimate at these radii and instead rely on
radially binned values taken from Drukier~\etal\ (1998, Col.~2 of
Table 4). A bias may occur at small radii as well, if both the density
of data and the value of the second moment change dramatically over a
small radial range.  We thus also estimate the second moment in a
central bin containing the inner ten stars.  The solid points in
Fig.~12 and their 68\% error bars are these binned values.

The value of the second moment for the central ten stars, with an
average radius~=~1\arcsec, is $11.7\pm2.8$~\kms.  The second moment
remains approximately constant out to a radius of 30\arcsec, a value
and trend identical to those found previously (Gebhardt~\etal\ 1997).

Gebhardt~\etal\ (1997) suggest that the projected rotation increases
within 3\arcsec\ of the center of M15 and that the position angle of
the projected rotation axis there differs by 100\deg\ from that at
larger radii.  This estimate came from the integrated cluster light,
as there were too few individual stellar velocity measurements to
constrain the rotation.  With the additional velocity information from
the AO dataset, we can measure the rotation from the stars alone. We
are interested in looking for significant changes in the rotation over
a small radial range. Smoothing techniques, such as the LOWESS
technique, tend to wash out small scale features. Although techniques
based on binning do not suffer from this particular effect, they are
sensitive to the subjective choice of bin width and bin location.  We
have tried to preserve the better features of both techniques by using
a hybrid approach: a variable-width data window in radius with lower
weights for the points near the window edges. The variable-width
window allows for recovery of sharp radial variations, and the
weighting scheme is designed to lessen possible large variations due
to discrepant datapoints.

We apply the data window to the velocities sorted into radial
order. The window contains 251 points, except that it shrinks to a
minimum of 25 points at the inner and the outer boundaries of the
data. At each position of the window a maximum likelihood estimator
determines the rotation amplitude, rotation position angle (PA), and
dispersion about the rotation. Within each bin, the weight of a point
varies as the cube of its distance from the bin center -- \ie\
$w_i=1-(|i-n/2|)^3/(n/2)^3$, where $n$ represents the number of points
in the subsample and $i$ is that point's rank in the sorted
dataset. This weight multiplies that point's contribution to the
likelihood function. The likelihood function minimizes the deviations
of the velocities minus the cluster systemic velocity from a mean
velocity that varies sinusoidally as a function of position angle. A
sinusoidal variation is expected for solid-body rotation. However, for
other rotation profiles, a sine function may not be optimal for
measuring the rotation amplitude. The radius corresponding to each
position of the window is the average of the linear radii of the
included points.

Fig.~13 plots the results from the hybrid technique for the following
four quantities, each with their 68\% confidence bands: the projected
rotation amplitude, the dispersion about the rotation, $v/\sigma$, and
the PA of the projected rotation axis.  Significant rotation exists at
most radii in the cluster; only at 0.3\arcmin\ is the net rotation
near zero, causing the PA estimate to fluctuate rapidly near that
radius.  Most surprising is the rotation seen at small radii.  Inside
of 0.05\arcmin, the rotation amplitude rises significantly and, at
0.04\arcmin, the rotational support equals the pressure support (\ie\
$v/\sigma=1$).  Fig.~14 displays this rotation directly by plotting
the offset from the mean cluster velocity for the 40 stars in the
central 3.4\arcsec\ vs.\ azimuth. The rotation amplitude for these
stars is $v = 10.4\pm2.7$~\kms\ and the dispersion after removing the
rotation is $\sigma = 10.3\pm1.4$~\kms, so $v/\sigma \simeq 1$ in this
region. The probability of measuring a rotation this high by chance
when none is present is much less than 1\%. Fig.~13 suggests that the
rotation amplitude remains high into the center.  Even for the five
stars inside of 1\arcsec, the rotation has the same position angle and
an amplitude equal to $15.3\pm 6.2$~\kms. For these five stars, a
rotation amplitude this large is expected by chance only 5\% of the
time.

The confidence bands in Fig~13 result from a Monte-Carlo analysis.  We
create 100 realizations of the velocity sample by drawing points from
the initial fit to the rotation as a function of radius. To each point
we add random uncertainties that come from the data uncertainty
distribution, determined by the velocity difference of the individual
measurements relative to the rotation profile; this procedure includes
both the velocity dispersion and the individual velocity
uncertainties.  To mimic the radial variation of dispersion and
uncertainty in the data (the velocities at large radii primarily come
from Drukier~\etal\ (1998), which are on average more precise than the
FP velocities), we have drawn the additive uncertainties from radial
bins. For each realization, we estimate the radial variation of the
kinematics as we did for the observed data. The scatter in the
distribution of the kinematical properties at each radius provide the
68\% confidence bands.

\vskip 60pt
\section{MASS MODELING}

Since our dataset is not dramatically larger than in our previous
analysis, we have not re-computed the mass density profile and the
stellar mass function.  Instead, we concentrate on the central
dynamics and, in particular, black-hole models, since the most
important changes in the dataset are at small radii.

Figure~15 plots both the actual second moment profile of the data from
Fig.~12 and the expected second moment profiles for models with no
rotation, an isotropic velocity dispersion tensor, a constant stellar
M/L, and black holes of various masses.  We construct the models from
the surface brightness profile used in Gebhardt~\etal\ (1997).  The
deprojection of that profile determines the potential, assuming a
constant M/L$_{\rm V}$ = 1.7.  We add black holes of various masses to
this potential and calculate the projected second moments using the
isotropic Jeans equation.  The model that best matches the data, given
these assumptions, contains a $2000~\msun$ black hole.  The full range
of acceptable black hole masses is 0 to $4000~\msun$.  Alternatively,
the data are consistent with models with no black hole that have a
stellar M/L$_{\rm V}$ that increases towards the center to a value of
5, possibly due to a central concentration of white dwarfs or neutron
stars.

An additional, and probably more interesting, constraint to the
dynamical models comes from our observed rotation profile.  The AO
data argue that net rotation exists inside of 3.4\arcsec\ (1.7
parsecs) with greater than 99\% confidence and the best estimate of
the rotation amplitude yields $v/\sigma = 1.0$.  The innermost 40
stars have an average radius of 2.1\arcsec\ and a projected rotation
amplitude of 10.4~\kms.  Assuming pure rotational support, this
amplitude implies a central mass concentration equal to $2500~\msun$.
Net rotation this large at these small radii is surprising, but might
be easier to explain with the presence of a few thousand solar mass
black hole.

\section{SUMMARY AND DISCUSSION}

Two of the most obvious and interesting features of the M15 kinematics
are the increase in the rotation amplitude at small radii, where
$v/\sigma \simeq 1$, and the difference in the rotation axis PA at
those same radii compared to the rotation of the rest of the cluster.
The cluster 47~Tuc also shows more rotation than expected at small
radii (Gebhardt~\etal\ 1994), though less dramatically than M15.
These two clusters have the largest velocity datasets presently
available, suggesting that large central rotation may be a common
feature among centrally-concentrated globular clusters.  Can these
features be understood in the context of dynamical evolution producing
a collapsed core?  Or does the rapid increase in net rotation suggest
the presence of a central massive black hole?  Unfortunately, few
theoretical studies of cluster dynamical evolution include net
rotation.  Studies of the effects produced by central black holes in
rotating systems are also rare.

\subsection{Rotation and Core Collapse}

The exchange of stellar energies and momenta caused by two-body
relaxation acts to eliminate gradients in the mean velocity and, thus,
to create solid-body rotation in a cluster with a net angular momentum
(e.g., Ogorodnikov 1965).  However, this equilibrium may not be
stable, so the real evolution of a cluster can be more complex.
Hachisu (1979) was one of the first to point out the gravo-gyro
instability, where the central regions shrink and then spin up to
maintain virial equilibrium as relaxation transports angular momentum
outwards.  This requires that the rotation supply significant support
against gravity.  N-body simulations of clusters with equal-mass stars
by Akiyama \& Sugimoto (1989) suggest that the random velocities
eventually increase more rapidly than the net rotation, so that
$v/\sigma$ decreases and the cluster again tends towards solid body
rotation.  Similar N-body simulations by Arabadjis \& Richstone
(1998a) found that the flattening (and, presumably, the $v/\sigma$) of
the central regions decreased more slowly when a spectrum of stellar
masses was present.  But it still decreased or, at best, remained
constant.

The N-body simulations were unable to follow the collapse of the core
very far because they only had N = 1000 and 3000, respectively.
Einsel \& Spurzem (1999) studied the dynamical evolution of a rotating
globular cluster by numerically solving the Fokker-Planck equation to
follow the change in the stellar distribution function.  The clusters
are tidally limited and contain stars of only a single mass.  These
simulations have much less noise than N-body models, though they are
also too simple for confident comparison with real clusters.  In their
simulations, the net rotation velocity in the core increases rapidly
at first, probably driven by the gravo-gyro instability, and then more
slowly, at the same rate that the central velocity dispersion
increases as the core collapses.  However, there is only a small
increase in $v/\sigma$ in the core and certainly no steep gradient in
this quantity near the center.  The amount of rotation at large radii
decreases, but the maximum mean rotation velocity always occurs at
about the half-mass radius. This result is consistent with our data
outside of the central few arcseconds, for which the maximum rotation
velocity occurs at 0.9\arcmin -- approximately the half-mass radius for
M15.

Preliminary multi-mass Fokker-Planck simulations (Einsel 1999) do show
a larger increase in the net rotation for the most massive stars in
the cluster center, perhaps similar to the results of Arabadjis \&
Richstone (1998a).  These models need further testing, however, and
are also being checked with additional N-body simulations (Spurzem,
private communication).  If the giants whose radial velocities we
measure are the heaviest stars in the cluster and have been
concentrated in the core by mass segregation, then we would expect the
central mass-to-light ratio to be lower than that of the rest of the
cluster.  Figure~7 of Gebhardt~\etal\ (1997) suggests that the
mass-to-light ratio increases at small radii, though the uncertainties
are large enough that a decrease can not be ruled out.

An argument against core collapse being the explanation for the
rotation at small radii in M15 is the large change in the position
angle of the apparent rotation axis.  None of the simulations
discussed above have any torque that could produce what is nearly
counter-rotation.  Tidal forces from our Galaxy can exert torques and
will act differently on the inner and outer regions of the cluster.
We would naively expect any resulting change in the rotation axis to
occur at much larger radii.  Perhaps gravitational coupling such as
that in warped disks could align the rotation over most of the
cluster, but the nearly spherical shape of M15 must make that weak and
we see no reason that small radii would be exempted. Two-body
relaxation also acts to align rotation throughout the cluster and
should be most effective at small radii, where the relaxation time is
short.

In summary, models of globular cluster dynamical evolution do not seem
to produce the changing rotation axis position angle and the
$v/\sigma$ profile with a sharp central upturn that we observe in M15.
The models might be more successful in explaining the more moderate
departure from solid-body rotation in 47~Tuc.  More simulations,
particularly with models containing a range of stellar masses are
needed, but we now turn to whether models with central black holes can
explain the M15 data.

\subsection{Rotation and a Central Black Hole}

Whether M15 contains a massive black-hole remains a central question
for black-hole demographics and has been the subject of much debate,
starting with the discovery of a central x-ray source (Giacconi~\etal\
1974, Bahcall~\& Ostriker 1975) and revived more recently by the
velocity dispersion measurements of Peterson~\etal\ (1989).  From
dynamical studies of galaxies, Kormendy \& Richstone (1995) and
Magorrian~\etal\ (1998) find a relation between the mass of the
black-hole and the mass of the host galaxy.  This relation is based on
massive stellar systems (M32 being the smallest) and we have not yet
explored its extrapolation to smaller systems. Possibly all hot
stellar systems harbor a central massive black hole; this situation
would have significant consequences in modeling these systems, and may
even suggest that black holes act as a seed for hot stellar systems.
Globular clusters are the natural targets to answer such a hypothesis
due to their low mass.  M15 is particularly well suited for these
studies due to its proximity (9 kpc) and the high surface brightness
of its cusp.
% xxx State what M_BH
%would be expected for M15 from the E galaxy relation?  Mention the
%differences between globulars and E's w.r.t. the timescale for
%collisions between stars?xxx

Bahcall \& Wolf (1976) calculated the mass density and velocity
dispersion profiles for the region of a non-rotating globular cluster
influenced by a central black hole.  The dispersion profile in Fig.~12
is consistent with their prediction, but we still do not have the
spatial resolution necessary to choose between the models presented
there.  Detailed analysis by Sosin \& King (1997) of the central
stellar cusp in M15 found that neither the black hole model of Bahcall
\& Wolf nor pure core-collapse models are a good match to the light
profile.  More sophisticated dynamical models, perhaps combining a
black hole with core collapse, are necessary.

Lee~\& Goodman (1989) demonstrated that the adiabatic growth of a
central black hole in a rotating stellar system increased the
projected $v/\sigma$ within the region of gravitational influence of
the black hole (where the mass of the stars approximately equals the
final mass of the black hole).  However, this study did not include
the effects of two-body relaxation.  The presence of the black hole
reduces the mass density of stars that would otherwise be deduced from
the velocity dispersion and this increases the relaxation time at
small radii, perhaps helping to maintain a larger rotation.  However,
this time remains much shorter than the age of M15 at small radii.
%xxxOr is it better to say that the increased dispersion increases the
%relaxation time (the stellar density presumably still increases
%inwards)?xxx

Arabadjis \& Richstone (1998b) followed the dynamical evolution due to
two-body relaxation of a system that initially had solid-body rotation
and stars of a single mass (the collisional coalescence of stars was
allowed) and contained a central black hole.  The rotation amplitude
and dispersion both increased near the center as the system evolved,
however, $v/\sigma$ was essentially constant in time.  The radial
profile of $v/\sigma$ at late times was flat.  These conclusions are
somewhat weakened by the noise in these N-body simulations (N = 3000),
which makes it hard to determine kinematic quantities over a large
range of radii.

Thus, while the models of Lee~\& Goodman (1989) are suggestive,
Fokker-Planck simulations including a black hole and net rotation are
probably needed to reach any definite conclusions.  A central black
hole with a mass of about $2500~\msun$ is consistent with all of the
dynamical data (we note that this mass is consistent with that from
Phinney's (1993) estimate of the enclosed mass from pulsar timing
measurements near the center of M15).  However, this model also
provides no clear explanation for the changing rotation axis position
angle.  Becoming (even) more speculative, stars formed in a disk of
gas, accreted from outside of the cluster, might explain the rotation
with or without a central black hole.  The color-magnitude diagram for
the inner region provides, at most, only a little evidence for a
younger stellar population at small radii (Guhathakurta~\etal\ 1996,
De Marchi \& Paresce 1995, Sosin \& King 1997).
% xxx did Raja et al.
% really never publish the cmds from their 1996 paper?

\subsection{The Potential of Adaptive Optics}

The high angular resolution provided by the Adaptive Optics Bonnette
on the CFHT have strengthened some of the most surprising results in
our previous M15 velocity samples.  While the increase in the sample
size was modest, the AO data provided crucial confirmation and, hence,
better velocity uncertainties for many of the stars closest to the
cluster center.  Even more importantly, the AO observations increased
the size of the velocity sample in this critical region. As adaptive
optics systems become better understood and available on larger
telescopes, and as the data analysis tools become better tailored to
these observations (\ie\ larger and better sampled PSFs), we will be
able to greatly improve our measurements of the kinematics near the
center of M15 and other core-collapse globular clusters.

Our AO observations have Strehl ratios of 2 -- 6\%.  Even under these
poor conditions for AO corrections, we were able to obtain photometry
with errors less than 2\%.  Of course, our observations were the best of
all worlds for photometry since we had bright stars everywhere in our
field. The ability to extend this detailed understanding of the
spatial and temporal variation of the PSF to observations where only a
handful of calibration stars are present is not clear.  Surely, present
and future AO studies will deal with this concern.

\acknowledgments

We are indebted to the CFHT staff who spent long hours setting up a
very difficult instrument configuration and who were able to provide
us with 0.09\arcsec\ FWHM images after 1 hour on the sky!  KG thanks
Jerry Nelson and Sandy Faber for many discussions about the AO
observations and reductions. KG is supported by NASA through Hubble
Fellowship grant HF-01090.01-97A awarded by the Space Telescope
Science Institute, which is operated by the Association of the
Universities for Research in Astronomy, Inc., for NASA under contract
NAS 5-26555.  The globular cluster research of TW and CP is supported
by grant AST96-19510 from the National Science Foundation.

\clearpage
\begin{figure}
\centerline{\psfig{file=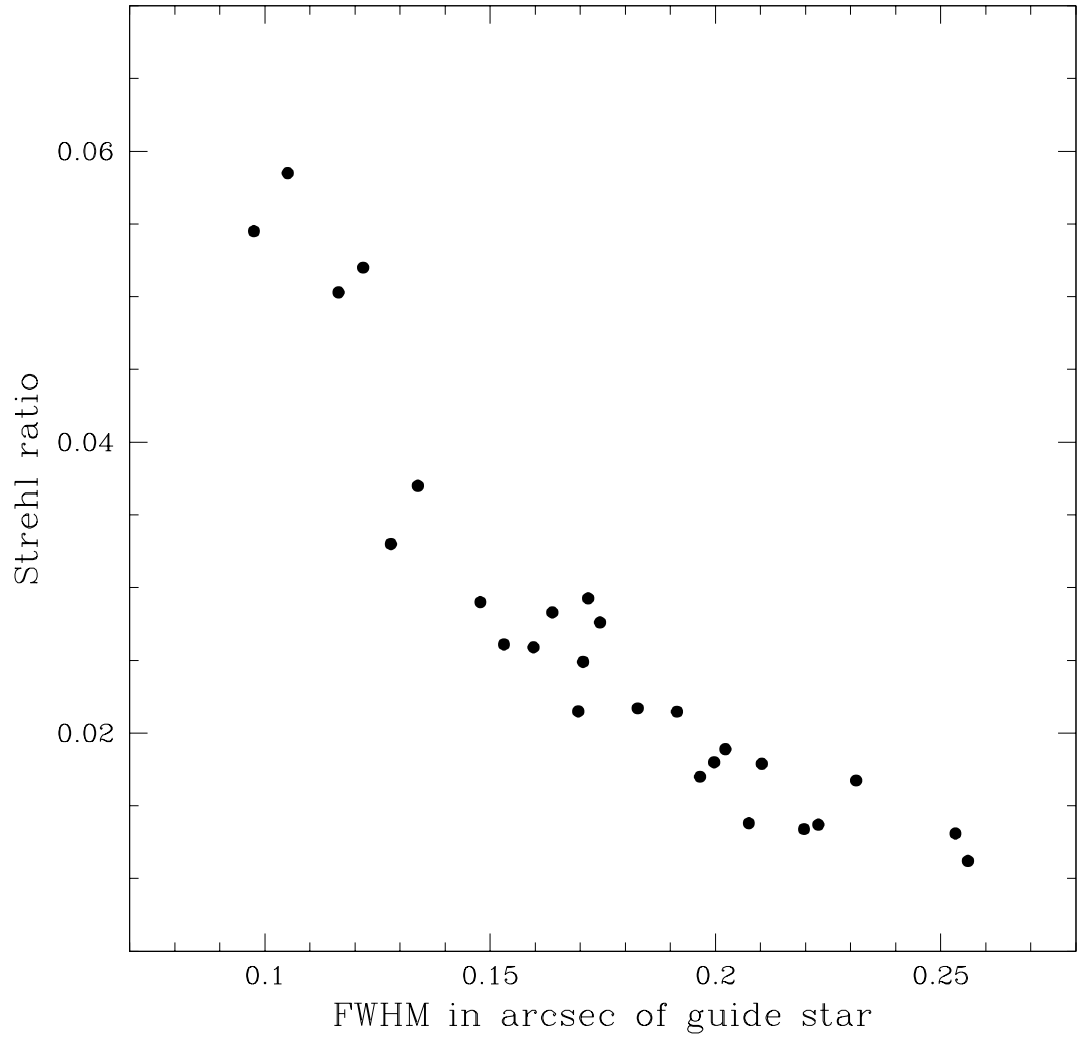,width=15cm,angle=0}}
%\plotfiddle{gebhardt.fig1.ps}{400}{0}{66}{66}{-210pt}{-20pt}
\caption{Strehl ratio of the guide star for each frame versus
the FWHM.}
\end{figure}

\clearpage
\begin{figure}
\centerline{\psfig{file=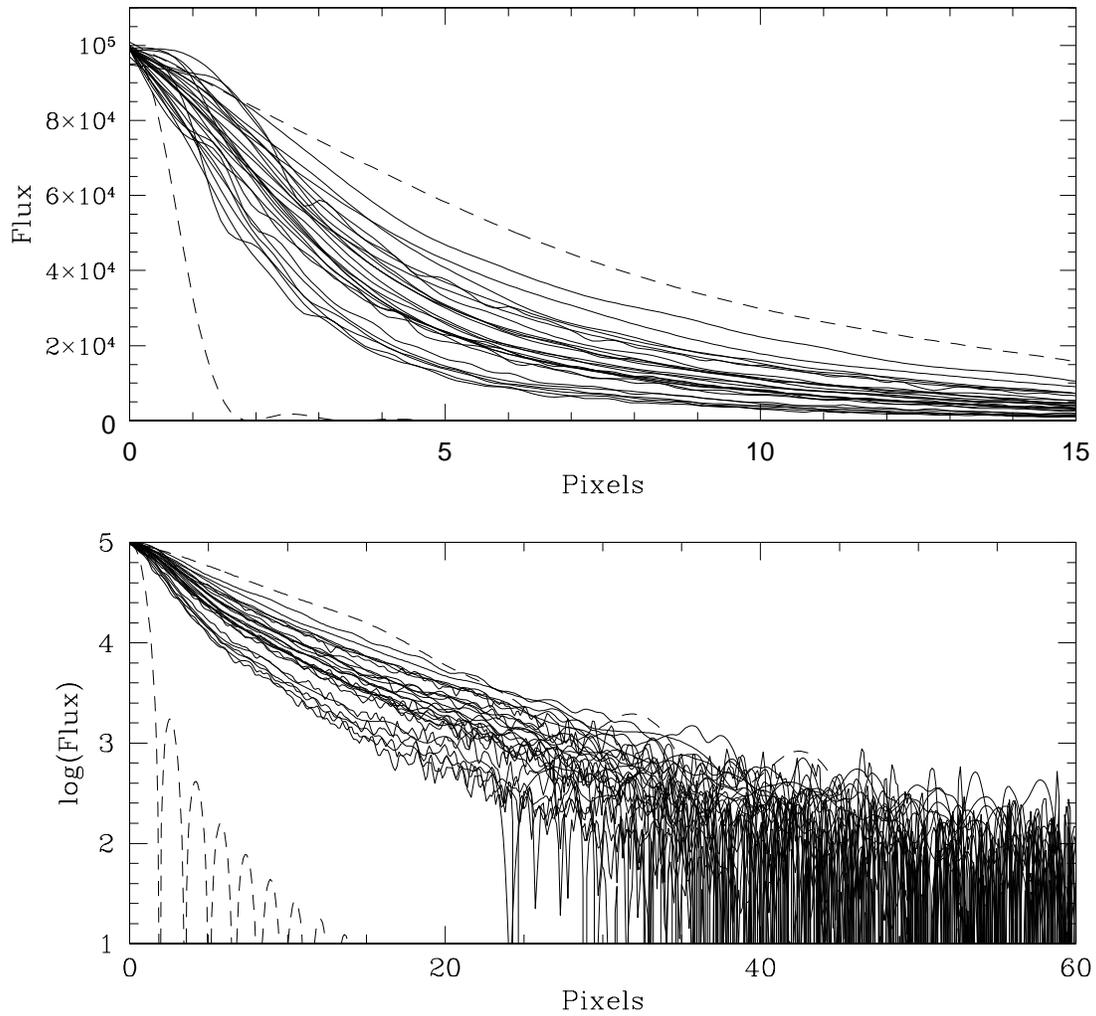,width=15cm,angle=0}}
%\plotfiddle{gebhardt.fig2.ps}{400pt}{0}{66}{66}{-210pt}{-20pt}
\caption{Flux of the guide star as a function of radius for each
frame. The upper plot is linear and the bottom logarithmic. The solid
lines represent the radial profile for the guide stars; the
upper-dashed line represents an uncorrected profile (\ie\ the AO
system was turned off), and the bottom dashed line represents a
diffraction-limited profile.}
\end{figure}

\clearpage
\begin{figure}
\centerline{\psfig{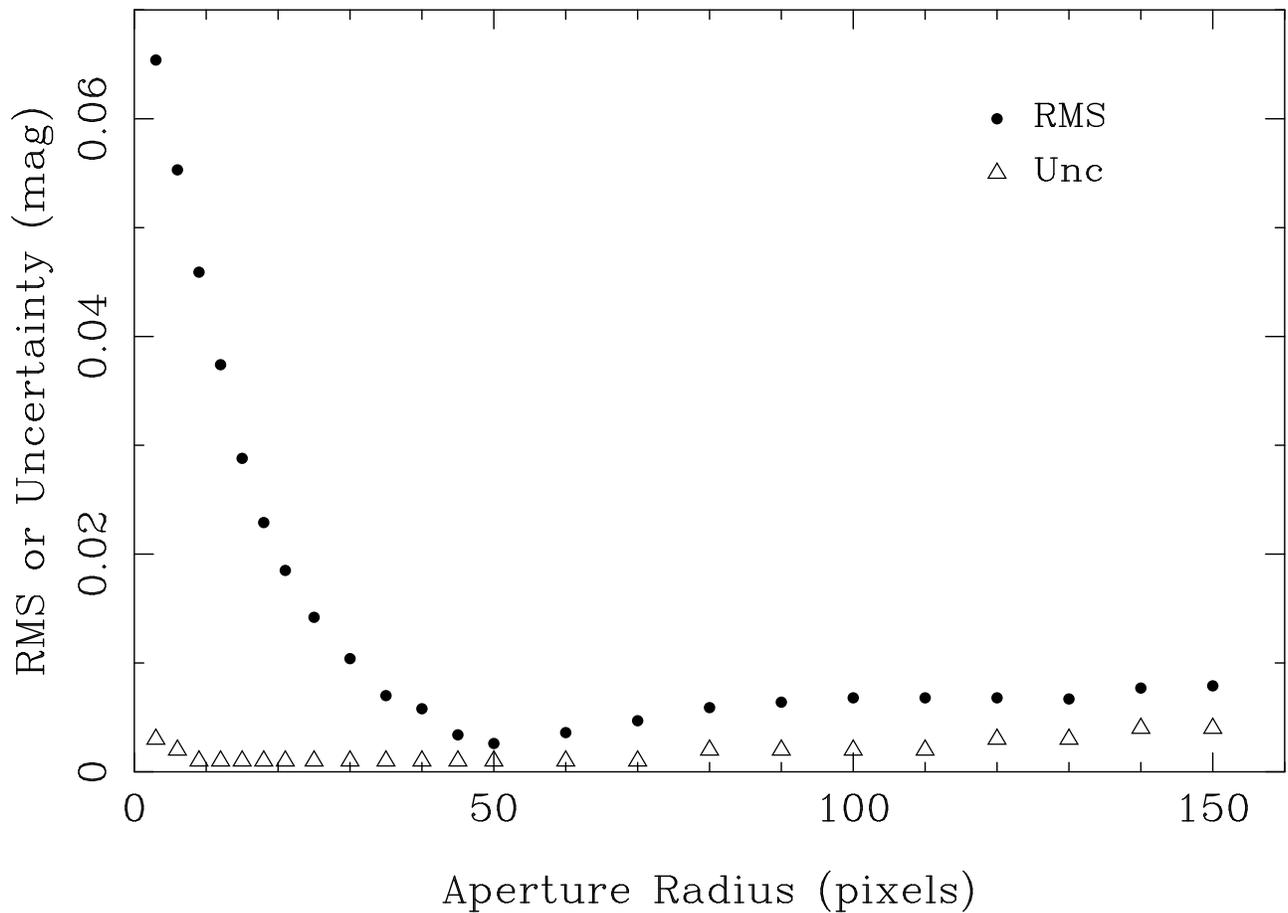}}
%\plotfiddle{gebhardt.fig3.ps}{350pt}{-90}{66}{66}{-250pt}{450pt}
\caption{The filled circles are the RMS scatter around the average
aperture magnitude for four consecutive 10~s exposures of HD107328.
The aperture radii vary between 3 and 150 pixels (1 pixel =
0.0305\arcsec).  The open triangles are the uncertainties in the
magnitudes expected from photon statistics.  Variations in the
magnitudes due to variable seeing and wavefront correction contribute
negligibly to the measurement uncertainties for apertures with radii
larger than 50 pixels (1.5 arcsec).}
\end{figure}

\clearpage
\begin{figure}
%\centerline{\psfig{file=gebhardt.fig4.ps,width=15cm,angle=0}}
%\plotfiddle{gebhardt.fig4.ps}{400pt}{0}{66}{66}{-210pt}{-20pt}
\centerline{\Large\bf 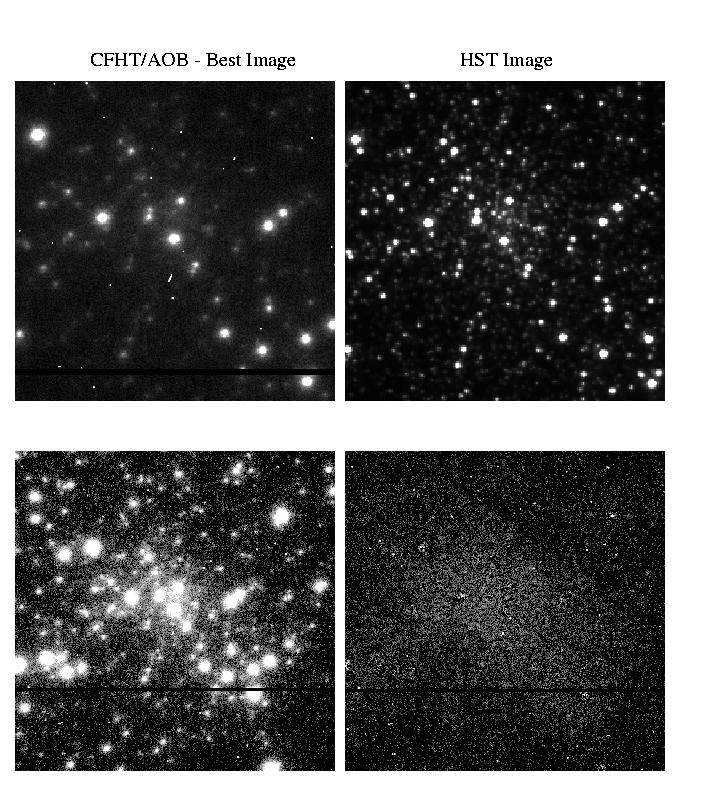 goes here}
\vskip 30pt
\caption{Four images of the central region in M15; the top two are the
central 9\arcsec\ square and the bottom two are the central 15\arcsec\
square (North is up and East is to the left). The top-left image is
the best-seeing CFHT/AOB frame with 0.09\arcsec\ FWHM displayed using
a linear stretch.  The top-right panel is an HST image taken from
Guhathakurta~\etal\ (1996).  The bottom-left image is a larger region
of the same CFHT image displayed with a square-root stretch to reveal
the fainter signal levels.  The bottom-right image is the
star-subtracted version of its neighbor to the left, again displayed
using a square-root stretch.}
\end{figure}

\clearpage
\begin{figure}
\centerline{\psfig{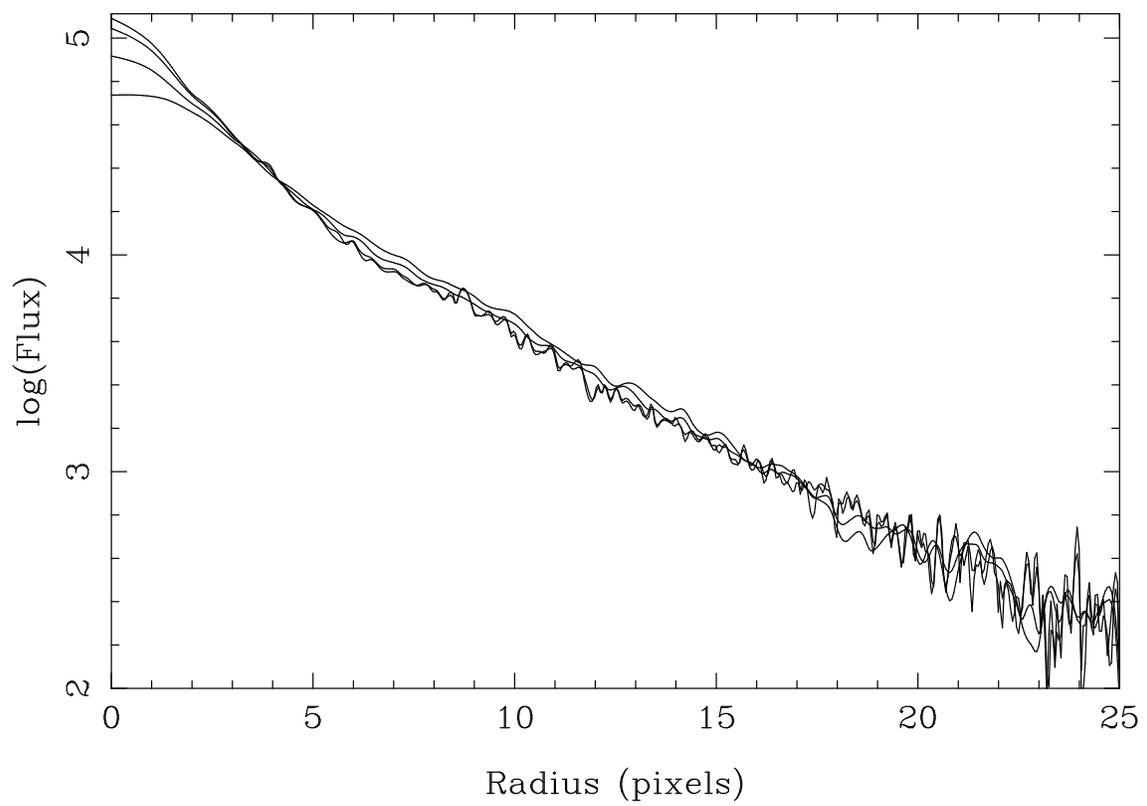}}
%\plotfiddle{gebhardt.fig5.ps}{350pt}{-90}{66}{66}{-250pt}{450pt}
\caption{Radial profile of the PSF at four different radial positions
in the frame.  The PSFs are normalized to have identical volume.}
\end{figure}

\clearpage
\begin{figure}
\centerline{\psfig{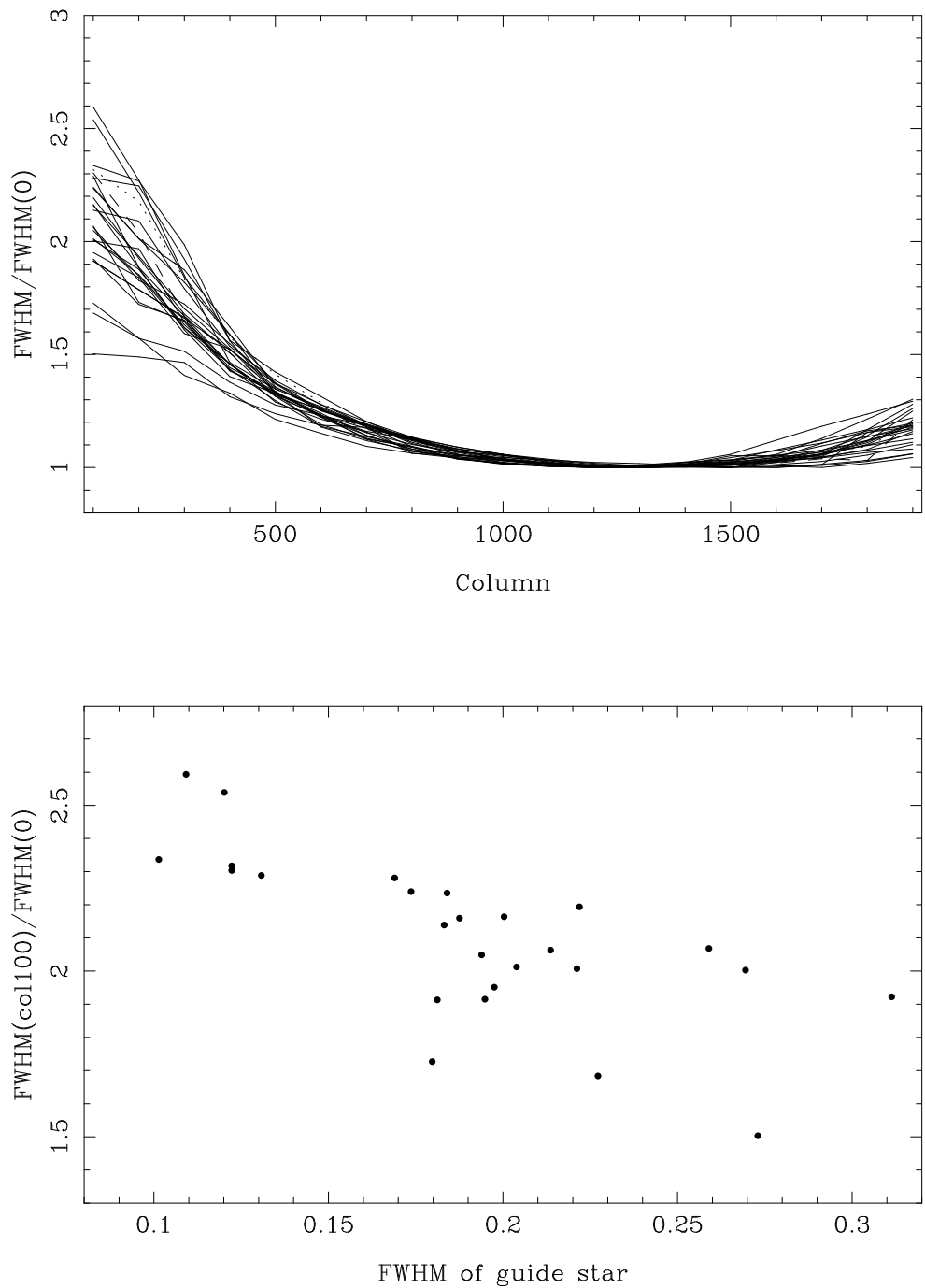}}
%\plotfiddle{gebhardt.fig6.ps}{400pt}{0}{66}{66}{-210pt}{-20pt}
\caption{The top panel shows the variation the PSF width with CCD
column number for the central row in the M15 frames.  Each line is the
the ratio of the FWHM of the stellar profile to the FWHM of the guide
star for a frame.  The bottom panel plots the FWHM ratio at the edge
of each image as a function of the guide star FWHM.  These points
suggest that the image with the best FWHM has the largest variation
across the field.}
\end{figure}

\clearpage
\begin{figure}
\centerline{\psfig{file=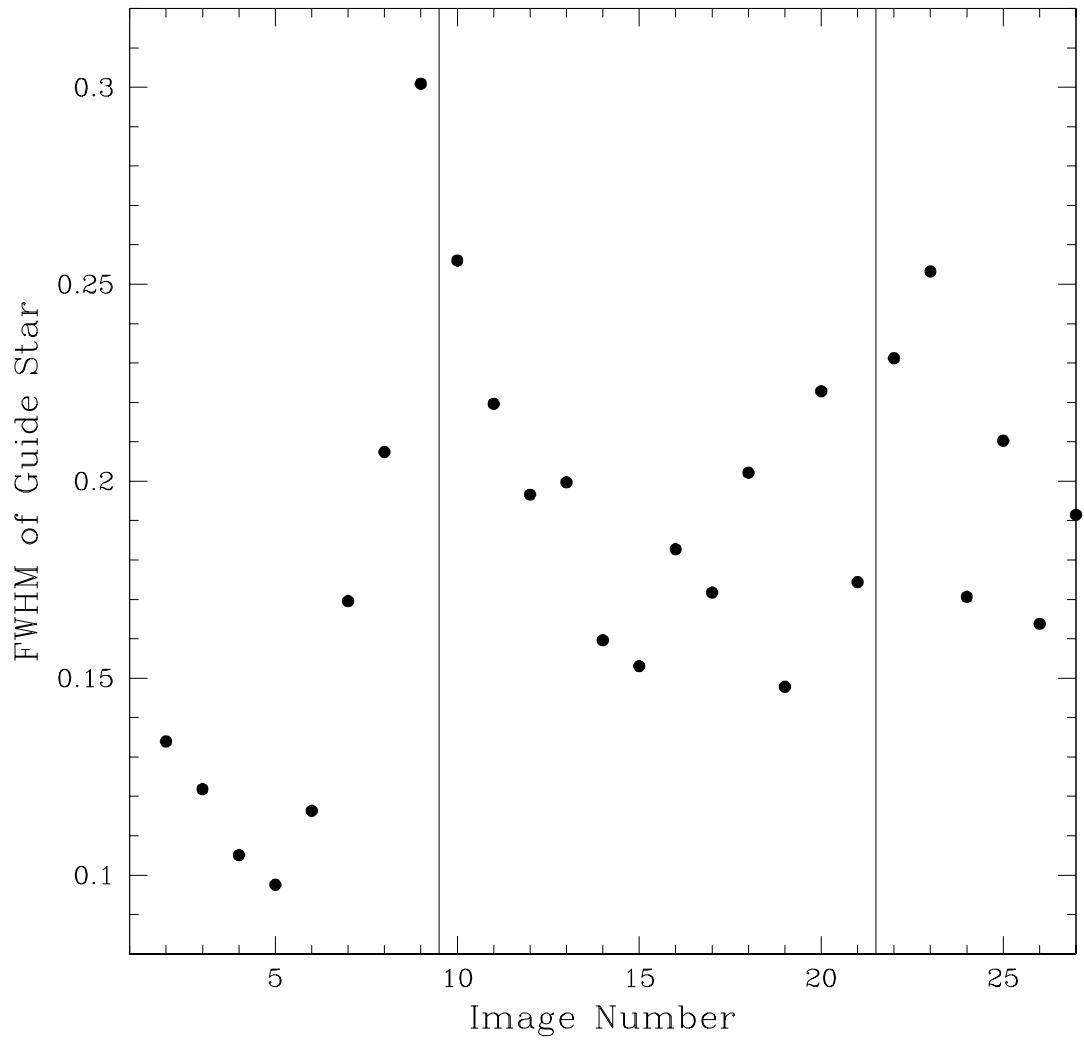,width=15cm,angle=0}}
%\plotfiddle{gebhardt.fig7.ps}{400pt}{0}{66}{66}{-210pt}{-20pt}
\caption{The temporal variation of the FWHM of the guide star from
frame to frame.  The vertical solid lines are the divisions between
the nights during which we took M15 data.}
\end{figure}

\clearpage
\begin{figure}
\centerline{\psfig{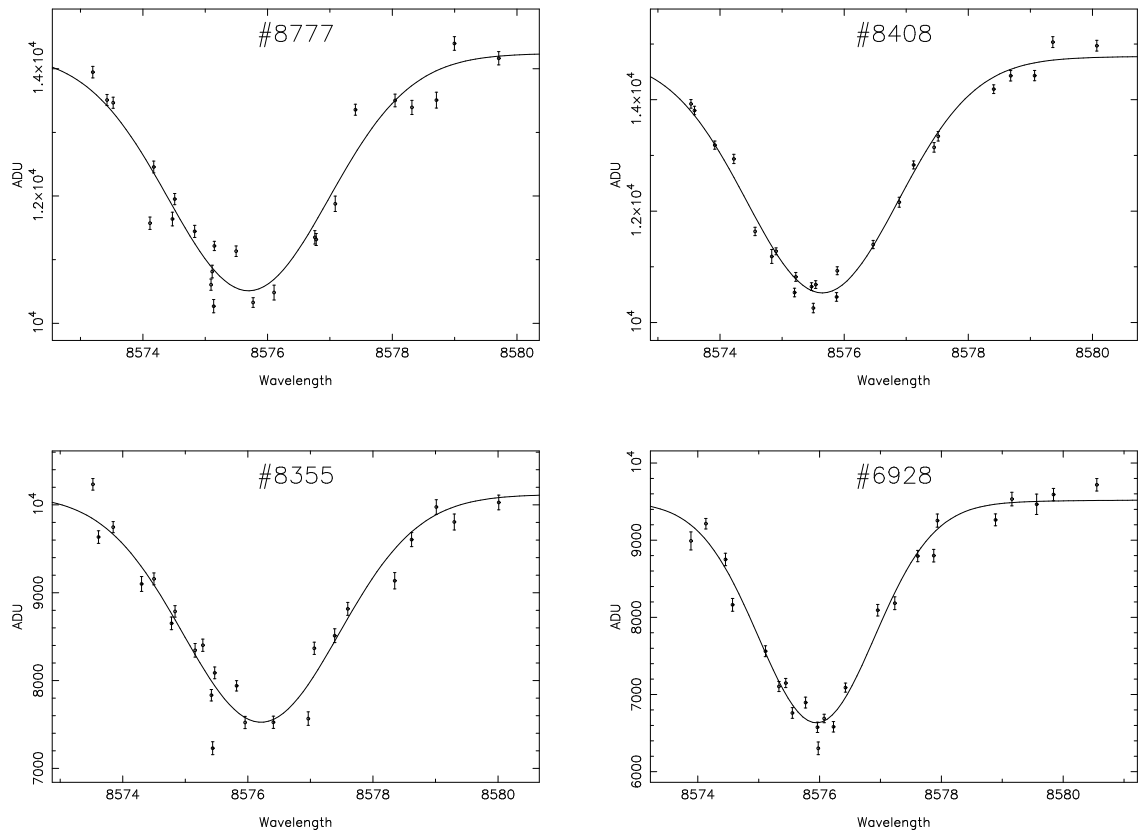}}
%\plotfiddle{gebhardt.fig8.ps}{350pt}{-90}{66}{66}{-250pt}{450pt}
\caption{Line profiles for four bright stars for which we were able to
obtain a reliable velocity measurement. The values and their error
bars come from ALLFRAME and include the frame normalizations. The
solid line is the fitted Voigt profile.}
\end{figure}

\clearpage
\begin{figure}
\centerline{\psfig{file=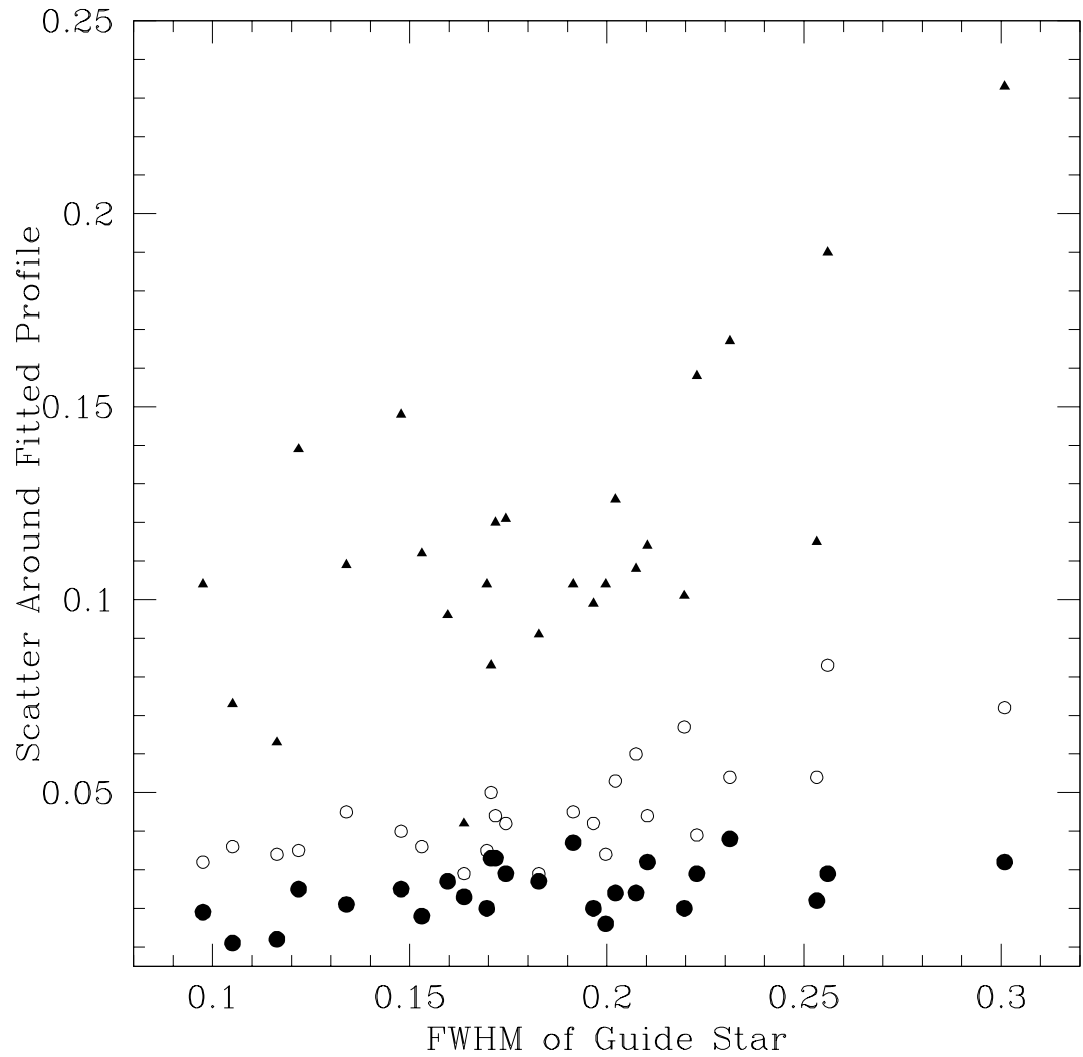,width=15cm,angle=0}}
%\plotfiddle{gebhardt.fig9.ps}{350pt}{0}{66}{66}{-210pt}{-20pt}
\caption{The dispersion in the fractional difference between the
fitted line profile and the actual spectral data for stars in each M15
image, plotted versus the FWHM of the guide star in the
image. Different symbols denote the dispersion for stars with
different brightnesses: solid circles are the brightest third, open
circles are the intermediate third, and solid triangles are the
faintest third.}
\end{figure}

\clearpage
\begin{figure}
\centerline{\psfig{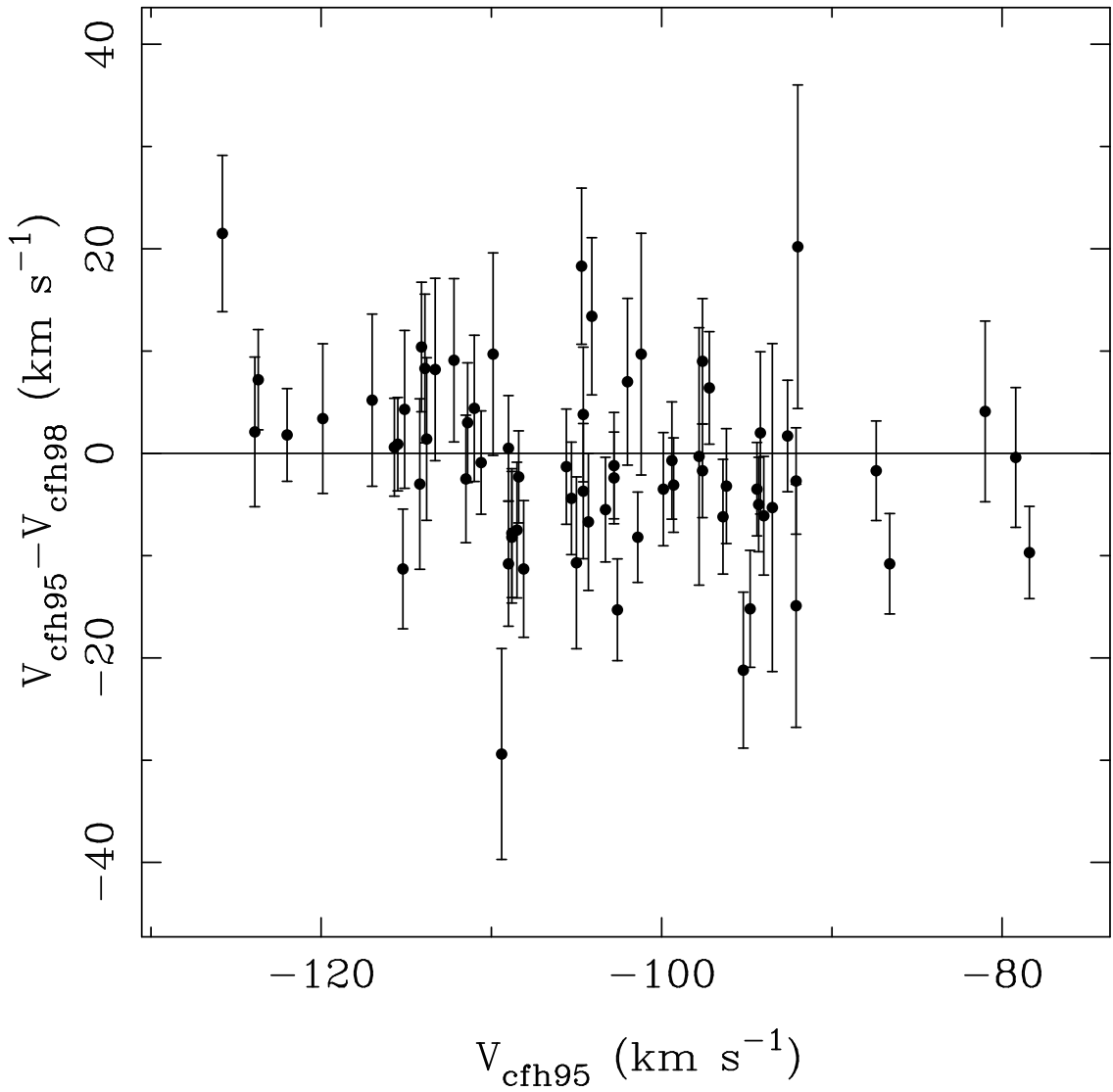}}
%\plotfiddle{gebhardt.fig10.ps}{350pt}{-90}{66}{66}{-250pt}{450pt}
\caption{A comparison of our velocities taken at CFHT in 1995 and the
velocities derived from the AO data.}
\end{figure}

\clearpage
\begin{figure}
\centerline{\psfig{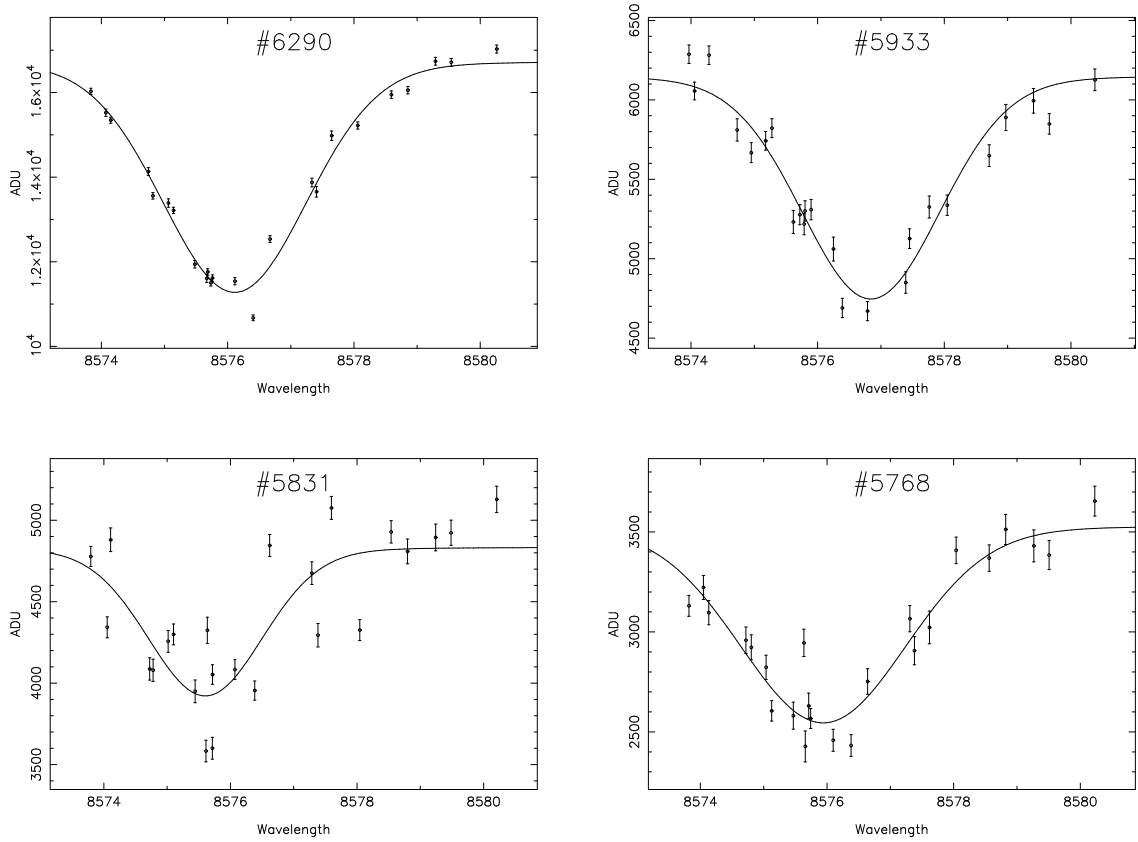}}
%\plotfiddle{gebhardt.fig11.ps}{350pt}{-90}{66}{66}{-250pt}{450pt}
\caption{Line profiles for the four innermost stars for which
we were able to obtain a reliable velocity measurement.}
\end{figure}

\clearpage
\begin{figure}
\centerline{\psfig{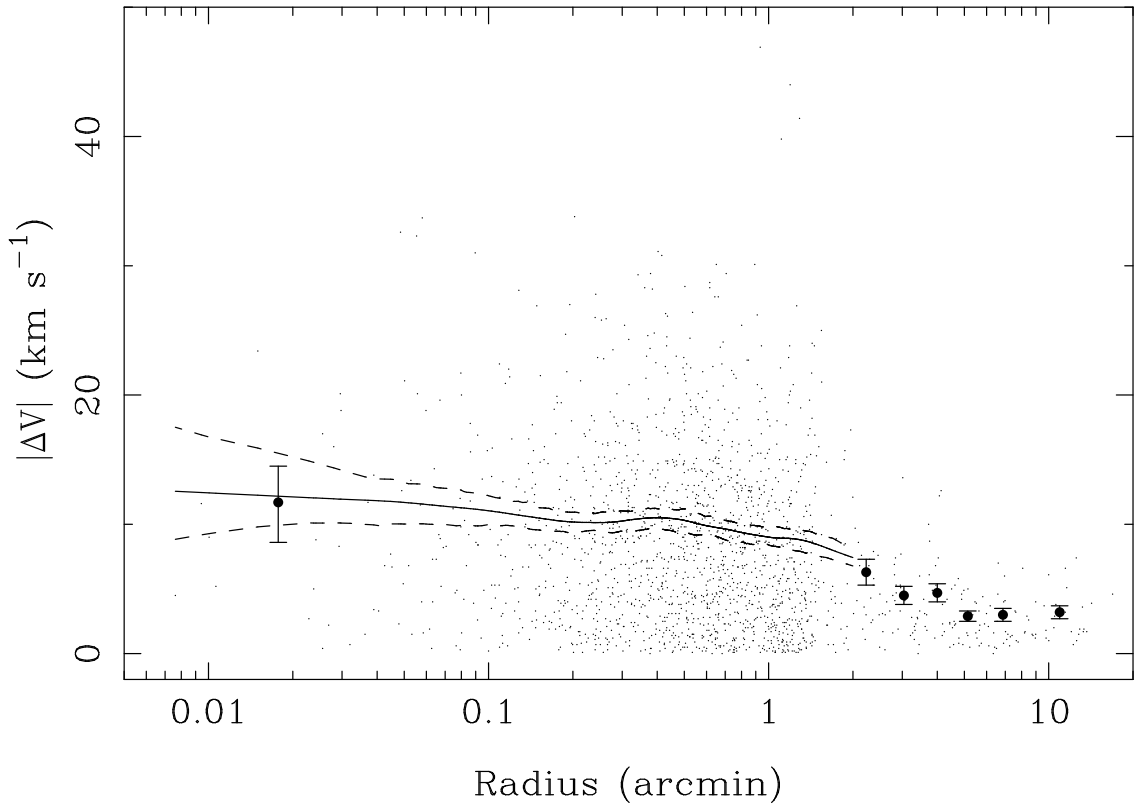}}
%\plotfiddle{gebhardt.fig12.ps}{350pt}{-90}{66}{66}{-250pt}{450pt}
\caption{The velocity dispersion profile for M15. The small dots are
the stellar velocity measurements. The solid and dashed lines
represent the velocity dispersion profile and the boundaries of its
90\% confidence band, respectively.  At large radii, the solid points
and their error bars are the binned dispersion estimates from
Drukier~\etal\ (1998).  The inner solid point represents the
dispersion measured from the innermost 10 stars of our sample.}
\end{figure}

\clearpage
\begin{figure}
\centerline{\psfig{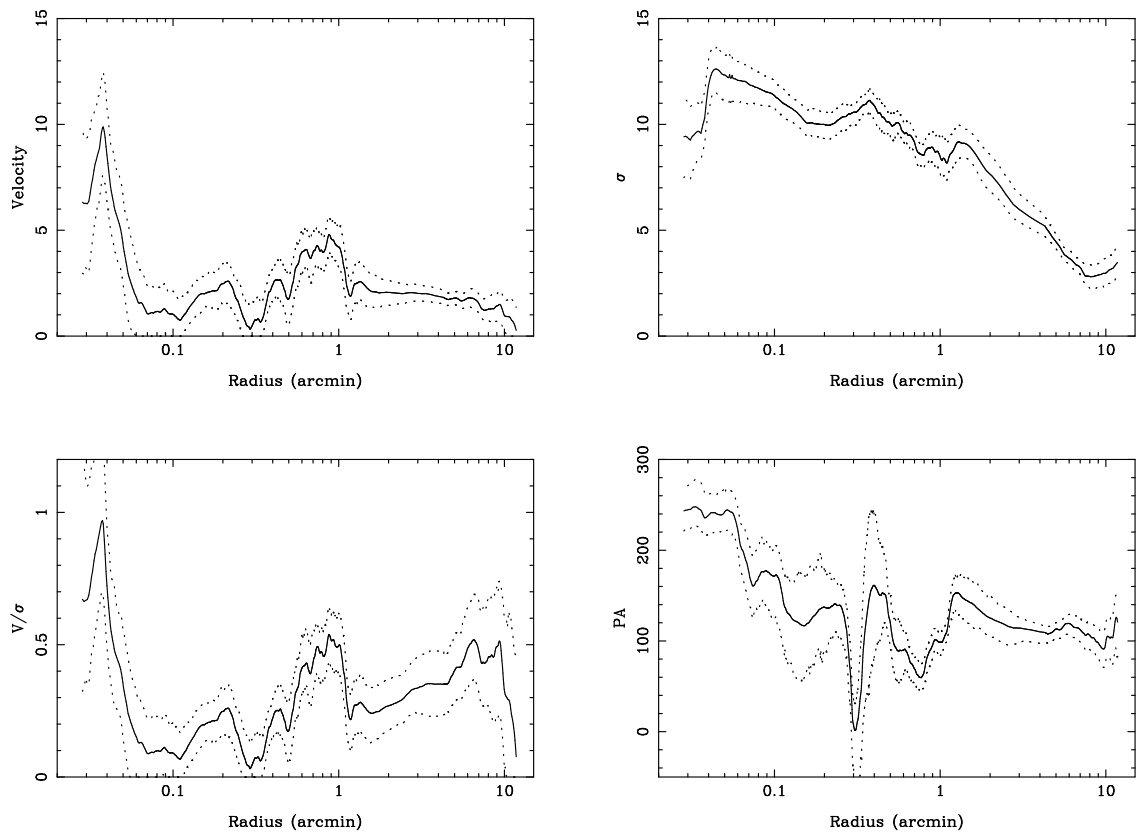}}
%\plotfiddle{gebhardt.fig13.ps}{350pt}{-90}{66}{66}{-250pt}{450pt}
\caption{The projected rotation amplitude, velocity dispersion,
$v/\sigma$, and the position angle of the projected rotation axis as
functions of radius.  The solid lines are the estimates and the dotted
lines are the boundaries of the 68\% confidence bands.}
\end{figure}

\clearpage
\begin{figure}
\centerline{\psfig{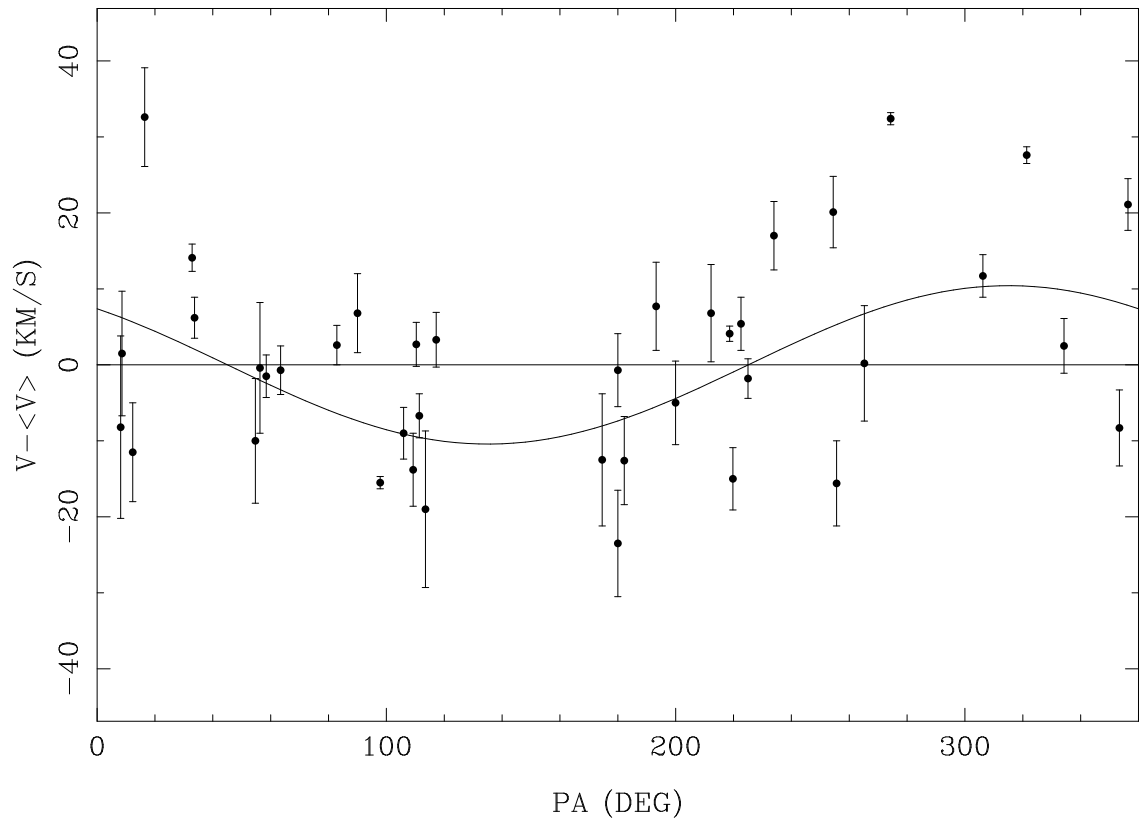}}
%\plotfiddle{gebhardt.fig14.ps}{350pt}{-90}{66}{66}{-250pt}{450pt}
\caption{The net projected rotation for the stars in the central
3.4\arcsec.  Each point is the offset of a star's velocity from the
cluster mean and its uncertainty.  The solid line is the best-fit
sinusoid with amplitude$= 10.8\pm2.6$ and PA$_0=226\pm14\deg$.  The
dispersion around this curve is 11.9~\kms.}
\end{figure}

\clearpage
\begin{figure}
\centerline{\psfig{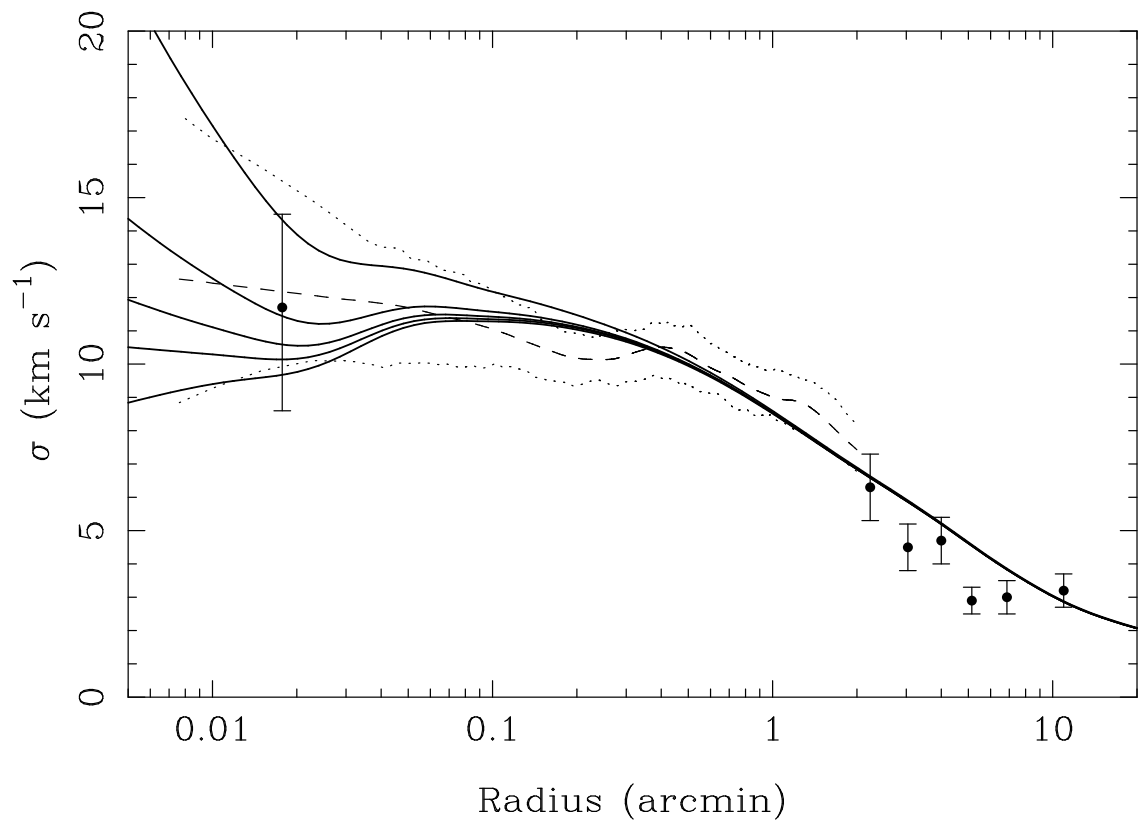}}
%\plotfiddle{gebhardt.fig15.ps}{350pt}{-90}{66}{66}{-250pt}{450pt}
\caption{The same dispersion profile as in Fig.~12 (the dashed line)
plotted with the profiles predicted by black hole models (the solid
lines).  The models assume isotropy, no rotation, and a constant
stellar M/L$_V$ of 1.7.  The masses of the black holes are equal to 0,
500, 1000, 2000, and 6000~$\msun$ in going from the lowest to the
highest profiles.}
\end{figure}


\begin{references}

\reference{} Akiyama, K., \& Sugimoto, D. 1989, PASJ, 41, 991
\reference{} Arabadjis, J.S., \& Richstone, D.O. 1998a, preprint 
astro-ph/9810192
\reference{} Arabadjis, J.S., \& Richstone, D.O. 1998b, preprint 
astro-ph/9810193
\reference{} Auri\`ere, M., \& Cordoni, J.-P. 1981, \aap, 138, 415
\reference{} Bahcall, J.N., \& Ostriker, J.P. 1975, Nature, 256, 23
\reference{} Bahcall, J.N., \& Wolf, R.A. 1976, \apj, 209, 214
\reference{} Beers, T.C., Flynn, K., \& Gebhardt, K. 1990, \aj, 100, 32
\reference{} Cohn, H. 1980, \apj, 242, 765
\reference{} De~Marchi, G., \& Paresce, F. 1995, \aap, 304, 202
\reference{} Djorgovski, S., \& King, I.R. 1986 \apjl, 305, L61
\reference{} Drukier, G., Slavin, S., Cohn, H., Lugger, P., Berrington, R.,
Murphy, B., \& Seitzer, P. 1998, \aj, 115, 708
\reference{} Dubath, P., \& Meylan, G. 1994, \aa, 290, 104
\reference{} Dubath, P., Meylan, G., \& Mayor M. 1995, \apj, 426, 192
\reference{} Dull, J.D. \etal\ 1997, \apj, 481, 267
\reference{} Einsel, C. \& Spurzem, R. 1999, \mnras, 302, 81
\reference{} Einsel, C. 1999, in ``Dynamics of Galaxies and Galactic 
Nuclei'' eds W.J. Duschl, C. Einsel, I.T.A. Series, No.2, Heidelberg, 
in press
\reference{} Esslinger, O., \& Edmunds, M. G. 1998, \aaps, 129, 617
\reference{} Gebhardt, K., Pryor, C., Williams, T.B., Hesser, J.E. 1994,
\aj, 107, 2067
\reference{} Gebhardt, K., Pryor, C., Williams, T.B., Hesser,
J.E. 1995, \aj, 110, 1699
\reference{} Gebhardt, K., Pryor, C., Williams, T.B., Hesser,
J.E., \& Stetson, P.B. 1997, \aj, 113, 1026
\reference{} Giacconi, R., Murray, S., Gursky, H., Kellogg, E., Schreier, 
E., Matilsky, T., Koch, D., \& Tananbaum, H. 1974, \apjs, 27,37
\reference{} Grabhorn, R. P., Cohn, H. N., Lugger, P. M., \& Murphy, B.
W. 1992, \apj, 392, 86
\reference{} Guhathakurta, P., Yanny, B., Schneider, D., \& Bahcall,
J. 1996, \aj, 111, 267
\reference{} Guhathakurta, P.~\etal\ 2000, in preparation.
\reference{} Gunn, J.E. \& Griffin, R. F. 1979, \aj, 84, 752
\reference{} Hachisu, I. 1979, PASJ, 31, 523
\reference{} King, I.R. 1975, in IAU Symposium No. 69, ``Dynamics of Stellar
Systems'', ed. A. Hayli (Dordrecht:Reidl), p.99
\reference{} Kormendy, J. \& Richstone, D. 1995, ARA\&A, 33, 581
\reference{} Lee, M.H., \& Goodman, J. 1989, \apj, 343, 594
\reference{} Magorrian, J. \etal\ 1998, AJ, 115, 2285
\reference{} Mayor, M. \etal\ 1983, \aaps, 54, 495
\reference{} Newell, B., Da~Costa, G.S., \& Norris, J. 1976, \apjl, 208, L55
\reference{} Ogorodnikov, K. F. 1965, Dynamics of Stellar Systems, 
translated from the Russian by J.~B.~Sykes (Oxford: Pergamon)
\reference{} Peterson, R.C., Seitzer, P., \& Cudworth, K.M. 1989, \apj,
347, 251 (PSC)
\reference{} Phinney, E.S. 1993, in Structure and Dynamics of Globular 
Clusters, eds. S.G. Djorgovski and G. Meylan, ASP Conf. Ser. No 50
(ASP, San Fransisco), p.141
\reference{} Rigaut, F., Salmon, D., Arsenault, R., Thomas, J., Lai, O.,
Rouan, D., V\'{e}ran, J. P., Gigan, P., Crampton, D., Fletcher, J. M.,
Stilburn, J., Boyer, C., \& Jagourel, P. 1998, \pasp, 110, 152
\reference{} Roberts, L. C., Jr., ten Brummelaar, T. A., \& Mason, B. D. 
1997, AAS Meeting 191, \#128.01
\reference{} Sosin, C., \& King, I.R. 1997, \aj, 113, 1328
\reference{} Spitzer, L. 1987, Dynamical Evolution of Globular Clusters
(Princeton: Princenton Univ.)
\reference{} Stetson, P.B. 1994, \pasp, 106, 250
\reference{} Stetson, P.B.~\etal\ 1998, \apj, 508, 491
\reference{} Yanny, B., Guhathakurta, P., Bahcall, J. \& Schneider, D.,
1994, \aj, 107, 1745
\reference{} Zaggia, S.R., Capaccioli, M., \& Piotto, G. 1993, \aap, 278,
415

\end{references}
\end{document}